\documentclass[12pt]{article}

\newcommand{\be}{\begin{equation}}
\newcommand{\ba}{\begin{eqnarray}}
\newcommand{\ea}{\end{eqnarray}}
\newcommand{\ee}{\end{equation}}

\newcommand{\beq}{\begin{equation}}
\newcommand{\eeq}{\end{equation}}
\newcommand{\beqa}{\begin{eqnarray}}
\newcommand{\eeqa}{\end{eqnarray}}

\newcommand{\unit}{\hbox to 3.8pt{\hskip1.3pt \vrule height 7.4pt
    width .4pt \hskip.7pt \vrule height 7.85pt width .4pt \kern-2.4pt
    \hrulefill \kern-3pt \raise 3.7pt\hbox{\char'40}}}

\begin{document}
\begin{titlepage}
\thispagestyle{empty}
\begin{flushright}
hep-th/0602271 \\
February, 2006
\end{flushright}

\bigskip

\begin{center}
\noindent{\Large \textbf{
Ghost D-brane, Supersymmetry and Matrix Model
}}\\
\vspace{2cm}
\noindent{
Seiji Terashima\footnote{seijit@physics.rutgers.edu} 
}\\

 {\it  New High Energy Theory Center, Rutgers University\\
126 Frelinghuysen Road, Piscataway, NJ 08854-8019, USA}
\vskip 2em
\bigskip

\end{center}
\begin{abstract}

In this note we study 
the world volume theory of pairs of D-brane and ghost D-brane,
which is shown to have 16 linear supersymmetries and
16 nonlinear supersymmetries.
In particular we study a matrix model based on the 
pairs of D$(-1)$-brane and ghost D$(-1)$-brane.
Since such pairs are supposed to be equivalent to the closed string vacuum,
we expect all 32 supersymmetries should be unbroken.
We show that the world volume theory of the pairs of D-brane and ghost D-brane
has unbroken 32 supersymmetries even though 
a half of them are nonlinearly realized.

\end{abstract}
\end{titlepage}

\newpage

%\tableofcontents

\section{Introduction}

As non-perturbative formulations of M-theory and string theory,
th BFSS matrix model \cite{BFSS} and the IKKT matrix model \cite{IKKT}
has been extensively studied.
Since the BFSS matrix model is supposed to describe 
M-theory in an infinitely boosted flame,
the action is 
the same as a low energy effective action on multiple BPS D$0$-branes in
type IIA superstring.
%A D$0$-brane in type IIA superstring has a conserved charge,
%i.e. D$0$-charge, 
From the conservation of the D$0$-charge, 
we can not construct D-branes without the
D$0$-charge in this action. 
This means that the
D-branes in the theory always need to have nonzero field
strengths and thus non-commutative world-volumes.
The IKKT matrix model has same problem if we think it is described by
the BPS D$(-1)$-branes in type IIB superstring.

In order to overcome this,
one might consider the matrix model based on
non-BPS D$0$-branes or ${\mbox{D}0-\overline{\mbox{D}0}}$ pairs
\cite{K-matrix}
since these branes have no conserved charges
and it was shown that
we can construct any D-branes from them \cite{Te, K-matrix}
using the boundary string field theory action \cite{KMM2} \cite{KL}
or boundary state.\footnote{
See \cite{Senreview} \cite{Senconjecture} \cite{SFT}
for the tachyon dynamics in open string theory.}
Furthermore, 
such unstable D-branes can decay into the closed string theory
then restore 32 supersymmetries.
In \cite{Yo}-\cite{Sen} such 32 unbroken supersymmetries in the action
of unstable D-branes were discussed.
On the other hand, the BPS D$0$-branes actions can 
have unbroken 16 supersymmetries only.
Thus it is interesting to study the matrix models based on the unstable D-branes.
However, the presence of the tachyons 
can not allow us to use a simple effective action 
for unstable D-branes
although the string field theory actions can be used at least 
in principle.\footnote{
For the two-dimensional string theory, 
it was proposed that the $c=1$ matrix model, which
had been known to describe the two-dimensional string theory,
can be considered as a tachyon action on
multiple D$0$-branes \cite{MV, KMS, MVT} and
in \cite{TaTe} it was indeed shown that
the boundary string field theory action for D$0$-branes 
is equivalent to the  $c=1$ matrix model for this case.
Maybe we can find some simple action for other cases.
}

Recently, a ghost D-brane in superstring theories
was introduced 
as an object that cancels the effects of a D-brane
\cite{OkTa}.
Thus a pair of D-brane and ghost D-brane at the same point is physically 
equivalent to the closed string vacuum.
This is similar to the pair of D-brane and anti-D-brane,
especially, after the tachyon condensation or VSFT \cite{VSFT}.
However, for the ghost D-brane case
we do not have the tachyon 
and then we can consider the ``low energy effective action'' 
for the D-branes and the ghost D-branes.
In particular, for the D0-brane or D$(-1)$-brane,
we can have a simple matrix model action for the pairs.
(Here the ghost D-brane has wrong sign for the kinetic term
and the spectrum contains fermionic scalars and bosonic spinors.
Thus the theory will be non-unitary for separated D-brane and ghost D-brane and
we should seriously consider a physical meaning of the low energy effective action
or the ghost D-brane itself,
although we will not do it in this paper.)

In this note we consider
this matrix model based on pairs of 
D$(-1)$-brane and ghost D$(-1)$-brane for type IIB superstring 
though a physical meaning of this matrix model is not clear by now.
Of course, we can consider 
D0-brane and ghost D0-brane for type IIA superstring, however,
we will concentrate on the D$(-1)$ brane action mainly for notational simplicity 
(Another reason is that 
a ghost D$(-1)$-brane action does not have kinetic term and
it might be easier to consider the (path-)integral for the action than 
other ghost D$p$-brane.)
Our main interest in this paper is 
supersymmetry on the matrix model.
Because pairs of D-brane and ghost D-brane without any nonzero vev 
will be equivalent to 
the closed string vacuum 
unlike the D-brane-anti-D-brane pairs, 
32 supersymmetries should be unbroken on it.
We will see that 32 supersymmetries of pairs of D-brane and ghost D-brane
are actually unbroken
despite the fact that a half of them are realized 
nonlinearly, which usually means the symmetries are broken.

The organization of this paper is as follows. 
In section two we study the world volume action of
pairs of D9-brane and ghost D9-brane.
In section three we propose a matrix model based on
D$(-1)$-brane and ghost D$(-1)$-brane and 
show the translational symmetry and 32 supersymmetries are unbroken.
In section four we draw conclusions
and discuss future problems.

\section{Pairs of D9-brane and ghost D9-brane}

In this section we consider the world volume theory on
$N$ pairs of D9-brane and ghost D9-brane
in type IIB superstring following \cite{OkTa}.\footnote{
The ghost D-brane was considered in \cite{ghost} and
some aspects of the ghost D-brane were considered implicitly in \cite{Vafa}.
}

It is well known that the low energy effective action of the 
$N$ D9-brane is the ten-dimensional $U(N)$ super Yang-Mills action,
\beq
L=-\frac{1}{4 g^2} {\rm tr}_{N \times N} 
\left( F_{\mu \nu} F^{\mu \nu} \right) 
-\frac{i}{2 g^2} {\rm tr}_{N \times N} 
\left( \bar{\lambda} \Gamma^\mu D_\mu \lambda \right),
\label{lym}
\eeq
where the gauge field $A_\mu$ and the gaugino $\lambda$,
which is a Majorana-Weyl spinor, are written in matrix notation
and the spinor index was not explicitly written.
The supersymmetry transformation is given by
\beqa
(\delta +\delta') A_\mu &=& -i \bar \zeta \Gamma_\mu \lambda \\
(\delta +\delta') \lambda &=& \frac{1}{2} F_{\mu \nu} \Gamma^{\mu \nu} \zeta + \zeta',
\label{susy}
\eeqa
where $\zeta$ corresponds to the unbroken 16 supersymmetries,
which is supposed to be linearly realized in the superfield formalism,
and $\zeta'$ corresponds to the nonlinearly realized 16 supersymmetries
which is broken by the presence of the D9-branes.

The world volume theory of $N$ pairs of D9-brane and ghost D9-brane
can be described by a gauge theory with $U(N|N)$ Chan-Paton matrices
and the low energy action in which massive fields and higher derivative terms are dropped
is given by the 
super Yang-Mills action with the supergroup $U(N|N)$ 
\cite{OkTa}.
The gauge field $A_\mu$ is replaced by 
\beq
\hat A_\mu =\left( 
\begin{array}{ll} 
A_\mu^{(1)}  & \chi_{\mu} \\
\chi^\dagger_{\mu}  & A_\mu^{(2)}  
\end{array} 
\right),
\eeq
where $A^{(i)}$ and $\chi$ are bosonic and 
fermionic $N \times N$ matrices, respectively.
$A^{(1)}$ (or  $A^{(2)}$) comes from the open string between the $N$ D-branes
(or the $N$ ghost D-branes) and
$\chi$ are from the open string between the D-branes and the ghost D-branes.
Similarly,
$\lambda$ is replaced by
\beq
\hat \lambda =\left( 
\begin{array}{ll} 
\lambda^{(1)}  & \varphi \\
\varphi^\dagger  & \lambda^{(2)}  
\end{array} 
\right),
\eeq
where $\lambda^{(i)}$ and $\varphi$ is 
a fermionic and bosonic spinor $N \times N$ matrices,
respectively.
Then the Lagrangian is given by
\beq
L=-\frac{1}{4 g^2} {\rm Str}_{2 N \times 2 N} 
\left( \hat F_{\mu \nu} \hat F^{\mu \nu} \right) 
-\frac{i}{2 g^2} {\rm Str}_{2N \times 2N} 
\left( \bar{\hat \lambda} \Gamma^\mu D_\mu \hat \lambda \right),
\label{lsym}
\eeq
where ${\rm Str}$ denotes the supertrace which is defined by
\beq
{\rm Str} \hat X={\rm tr} A - {\rm tr} D,
\eeq
where 
\beq
\hat X=\left( 
\begin{array}{ll} 
A & B \\
C  & D  
\end{array} 
\right).
\eeq

It may be important to note that the gauge group $U(N|N)$ does not have 
decoupled $U(1)$ part unlike (\ref{lym}) and
it may be significantly different from the action with gauge group $SU(N|N)$.\footnote{
Similar phenomena happen in noncommutative and non-anti-commutative gauge theory
with $U(N)$ gauge group \cite{SeWi}.
For the superalgebra, see \cite{Kac}.
}
This fact can be seen as follows.\footnote{
This $U(1)$ part in $U(N|N)$ was discussed in \cite{UNN} more carefully.}
The gauge field $\hat A_\mu$ can be written as
\beq
\hat A= \frac{1}{2N}  A^{tr} \unit_{2 N} 
+\frac{1}{2N}  A^{Str} K+ A^a T_a,
\eeq
where 
\beq
K=\left( 
\begin{tabular}{ll} 
$\unit_N$ & 0 \\
0 & $-\unit_N $
\end{tabular} 
\right)
\eeq
and $T_a$ which is bosonic and satisfies ${\rm Str} T_a={\rm tr} T_a=0$ 
and $ \unit_{2 N} $ form the generators of 
the $SU(N|N)$ subgroup of the $U(N|N)$.
Then the kinetic term containing $A^{tr}$ and/or $A^{Str}$ is proportional to
$(\partial_\mu A_\nu^{tr} -\partial_\nu A_\mu^{tr})
(\partial_\mu A_\nu^{Str} -\partial_\nu A_\mu^{Str})$
because of ${\rm Str} (\unit_{2 N} K)=2 N $ and 
${\rm Str} (\unit_{2 N} \unit_{2 N} )={\rm Str} (K K)={\rm Str} (\unit_{2 N} T_a)=
{\rm Str} (K T_a)=0$.
There is no interaction term for $A^{tr}$
since interaction terms are written by commutators, however,
it is easy to see there are interaction terms contain
$A^{Str}$ and $A^a$.
Therefore both $A^{Str}$ and $A^a$ are not decoupled from others.
Note that for the $N$ D-brane and $M$ ghost D-branes with $N \neq M$
the gauge group $U(N|M)$ contains a decoupled $U(1)$ part and 
we can decompose $U(N|M)=U(1) \times SU(N|M)$.

Supersymmetry transformations for the Lagrangian (\ref{lsym})
can be easily obtained from (\ref{susy}) since only the gauge group was changed.
Only a problem is that the fermion $\zeta$ does not 
commute with $\hat A_\mu$ and such property is need to show the invariance.
This is because the supermatrix contain fermions in off-diagonal parts.
However,
if we introduce
\beq
\hat \zeta = \zeta K,
\eeq
we find, for example,  $\zeta \hat A_\mu=(K \hat A_\mu K) \zeta$
and then $[\hat A_\mu, \hat \zeta]=0$. \footnote{
The constant grasmman variable was considered in
\cite{Park} in a different context.}
We also find that $\{ \hat \lambda, \hat \zeta \} =0$.
Hence the supersymmetry transformations for (\ref{lsym}) are given by 
\beqa
(\delta +\delta') \hat A_\mu &=& -i \bar{ \hat \zeta} \Gamma_\mu \hat \lambda \\
(\delta +\delta') \hat \lambda &=& \frac{1}{2} \hat F_{\mu \nu} \Gamma^{\mu \nu} \hat \zeta 
+ \hat \zeta'.
\label{ssusy}
\eeqa
Actually, if we formally expand ``bosonic'' superfield $\hat A$ as $\hat A=A^a \hat T_a$, 
where $A^a$ is a usual bosonic field and $\hat T_a$ is the ``bosonic'' supermatrix,
and ``fermionic'' superfield $\hat \lambda$ as $\hat \lambda=(\lambda^a K) \hat T_a$,
where $\lambda^a$ is a usual fermionic field,
then ${\rm Str}_{2N \times 2N} 
\left( \bar{\hat \lambda} \Gamma^\mu D_\mu \hat \lambda \right)$ in (\ref{lsym})
contains ${\rm Str}_{2N \times 2N} 
\left(\hat T_a [\hat T_b, \hat T_c ]  \right)=f_{abc}$
where $f_{abc}$ is bosonic constant antisymmetric for $a,b,c$.
%The paramereters $A^a$ and $\lambda^a$ are regarded as a usual boson and a fermion,
%respectively.
Thus, using this basis 
we can trivially follow the standard computation showing the invariance 
of (\ref{lym}) under the supersymmetry.

Here the supersymmetry transformations associated with $\hat \zeta'$
are nonlinear and they seem to be broken.
This seems to contradict the fact that the pair of D-brane and ghost D-brane is 
equivalent to the closed string vacuum.
However, in the next section, we will see nonlinearly realized symmetries can be
unbroken.

\section{Matrix Model Based on D-brane and ghost D-brane}

Now we consider $N$ pairs of D$(-1)$-brane and ghost D$(-1)$-brane
and the low energy effective action of those.
The action is given by 
the dimensional reduction of (\ref{lsym}) to 0 dimension
which replace $\hat A_\mu(x)$ (and $\hat \lambda(x)$) 
to $\hat \Phi_\mu$ (and $\hat \psi$):
\beq
S=-\frac{1}{4 g^2} {\rm Str}_{2 N \times 2 N} 
\left( \left[ \hat \Phi_\mu,  \hat \Phi_\nu \right] 
\left[ \hat \Phi^\mu,  \hat \Phi^\nu \right]\right) 
-\frac{1}{2 g^2} {\rm Str}_{2N \times 2N} 
\left( \bar{\hat \psi} \Gamma^\mu \left[ \Phi_\mu, \hat \psi \right]  \right),
\label{sm}
\eeq
where
\beq
\hat \Phi_\mu =\left( 
\begin{array}{ll} 
\Phi_\mu^{(1)}  & \chi_{\mu} \\
\chi^\dagger_{\mu}  & \Phi_\mu^{(2)}  
\end{array} 
\right)
\eeq
and
\beq
\hat \psi =\left( 
\begin{array}{ll} 
\psi^{(1)}  & \varphi \\
\varphi^\dagger  & \psi^{(2)}  
\end{array} 
\right).
\eeq
The supersymmetry transformations are
\beqa
(\delta +\delta') \hat \Phi_\mu &=& i \bar{ \hat \zeta} \Gamma_\mu \hat \phi \\
(\delta +\delta') \hat \psi_\mu &=& \frac{i}{2} \left[ \hat \Phi_\mu, \hat \Phi_\nu \right]  
\Gamma^{\mu \nu} \hat \zeta 
+ \hat \zeta'.
\label{susym}
\eeqa
Of course, this matrix model is also obtained from the IKKT matrix model action,
which is the 0 dimensional reduction of the 
10 dimensional $U(N)$ super Yang-Mills theory, 
by replacing $U(N)$ by supergroup $U(N|N)$.\footnote{
For other supermatrix models, see \cite{SM}-\cite{Morris}.
Note that the action (\ref{susym}) is not only a supermatrix model, 
i.e. supergauge symmetric,
but also supersymmetric.
Thus this is a supersymmetrized supermatrix model.
}
Note that a $U(1)$ factor is decoupled from other generators
of $U(N|N)$ for this matrix model
because there is no kinetic term, though $U(N|N)/U(1)$ is not 
the $SU(N|N)$ subgroup.

As the IKKT matrix model has been proposed as a nonperturbative formulation 
of type IIB superstring,
this matrix model could give a nonperturbative formulation 
of type IIB superstring in some large $N$ limit.
It is very interesting to investigate this possibility, however,
we will only consider symmetry of it here.

There is a constant shift symmetry,
\beq
\delta \hat \Phi^\mu = a^\mu, \;\;\; \delta \hat \psi=0,
\eeq
which can be understood as the space-time translation
of the D$(-1)$-brane and the ghost D$(-1)$-brane.
This symmetry is realized nonlinearly and 
if we separate the D-brane and ghost D-brane it will be broken 
by the presence of the D-brane.
However, for $\hat \Phi=\hat \psi=0$ 
it should be unbroken
since both the D-brane and the ghost D-brane disappear.\footnote{
Strictly speaking, there are no symmetry breaking in 0 or 1 dimensional theories.
We could rigorously define broken or unbroken symmetry 
by considering higher dimensional analogues,  
compactified theory or a large $N$ limit.}

To resolve this puzzle,
we first consider vev of a possible order parameter classically.
It should be gauge invariant and then 
$\langle {\rm Str} (\delta \hat \Phi_\mu) \rangle$
or the supertrace of polynomials of $\hat \Phi, \hat \psi$ 
can be considered.
It is easy to see that
\beq
\langle {\rm Str} (\delta \hat \Phi_\mu) \rangle={\rm Str} (a_\mu)=0,
\eeq
and the transformation of other gauge invariant operators also vanish
for $\hat \Phi=\hat \psi=0$ 
classically.
In this way, the nonlinear transformation 
can be regarded as unbroken.
Furthermore, we expect it is unbroken quantum mechanically.
Actually for $U(N|N)$ all correlation functions of gauge invariants
will vanish at $\hat \Phi=\hat \psi=0$ \cite{SM} \cite{OkTa} .
This is consistent with the interpretation as a closed string vacuum.
The transformation of the correlation functions also vanish
because the $\delta$ transformed $\hat \Phi$ to just a constant.
Therefore it is unbroken quantum mechanically.

On the other hand, for a generic classical background
\beq
\hat \Phi={\rm diag} (b_1, b_2, \ldots, b_N, c_1,c_2, \ldots, c_N), 
\;\;\; \hat \psi=0,
\label{bk}
\eeq
it will be broken.
Let us consider a gauge invariant 
$ {\rm Str} ( f_1(\hat \Phi, \hat \psi)) $
and the transformation of it,
$\delta {\rm Str} (f_1(\hat \Phi, \hat \psi) )$,
where $f_i$ is some polynomial.
Taking $f_1=\hat \Phi^{\mu_1} \cdots \hat \Phi^{\mu_M}$,
we have
\beq
\delta {\rm Str} (f_1(\hat \Phi, \hat \psi) )=
a^{\mu_1} \sum_{i=2}^N  \left( \prod_{a=1}^M (b^{\mu_a}_i)^2 
-\prod_{a=2}^M (c^{\mu_a}_i)^2 \right) 
+({ permutation \, of \, \mu_1 \, and \, \mu_i}).
\eeq
Therefore if 
\beq
b_i^\mu=c_i^\mu, \;\; (i=1, \ldots, N, \;\; \mu=0, \ldots, 9)
\label{cond}
\eeq
(or that with a permutation of $N$ vectors $c_i$) is satisfied,
$\delta {\rm Str} (f_1(\hat \Phi, \hat \psi) )=0$.
Considering the transformation of
$\langle {\rm Str} ( f_1(\hat \Phi, \hat \psi)) 
{\rm Str} ( f_2(\hat \Phi, \hat \psi)) \cdots \rangle$,
we see that
the translational symmetry is unbroken if (\ref{cond}) is satisfied
otherwise it is broken.
This is consistent with the interpretation that
a D-brane and a ghost D-brane in any pair
are at same position for (\ref{cond}) and 
are physically equivalent to the closed string vacuum.

We expect the nonlinear supersymmetries generated by $\hat \zeta'$
is also unbroken for (\ref{cond}) in the same way.
Here we note that 
${\rm Str} \hat \psi$ is not gauge invariant
because ${\rm Str} ( \hat X  \hat \psi ) \neq 
{\rm Str} (  \hat \psi X)$ where $\hat X$ is a usual supermatrix
and $\psi$ is a ``fermionic'' supermatrix, i.e. a 
supermatrix with fermions in its diagonal part.
Instead,
${\rm Str} (K \hat \psi)={\rm tr} (\hat \psi)$ is gauge invariant
because ${\rm Str} (K \hat X  \hat \psi ) = 
{\rm Str} (K  \hat \psi X)$ and 
consistent with the fact that a ``fermion'' bilinear 
${\rm Str(\hat \zeta' \hat \psi)}$ is gauge invariant.\footnote{
We also note that 
${\rm Str} ( \hat \psi  \hat \psi' ) =-{\rm Str} (  \hat \psi' \hat \psi)$
which is same property as the trace for usual matrices.
}
Then, it is easy to see that
\beq
\langle {\rm Str} (K \delta \hat \psi) \rangle={\rm Str} (K \hat \zeta')=
\zeta' {\rm Str} (K^2)=0,
\eeq
and the supersymmetric transformation of other gauge invariant operators also vanish
for $\hat \Phi=\hat \psi=0$ 
classically.
For the generic classical background (\ref{bk})
we can easily show that
(\ref{cond}) is the condition 
for the nonlinear supersymmetries being unbroken 
by taking $f_1=\hat \Phi^{\mu_1} \cdots \hat \Phi^{\mu_M} \hat \psi$.
Therefore the matrix model (\ref{sm}) has the 32 unbroken supersymmetries and 
10 dimensional translation symmetry (and the Lorentz symmetry linearly realized).
It is very interesting to investigate this highly symmetric matrix model further.
%In Appendix A, we will disucss the superalgebra.

Finally, 
we will discuss the supersymmetry algebra.
In \cite{IKKT} it was shown that the IKKT matrix model has 
the 32 supersymmetries which form the super symmetry algebra. 
We can trivially extend 
it to our case.
Indeed, if we define 
$\delta^{(1)}_{\hat \zeta}=\delta_{\hat \zeta}+\delta'_{\hat \zeta}$ and 
$\delta^{(2)}_{\hat \zeta}=i(\delta_{\hat \zeta}-\delta'_{\hat \zeta})$,
then we have the algebra of 32 supersymmetries 
\beq
[\delta^{(i)}_{\hat \zeta},\delta^{(j)}_{\hat \xi}]
=\epsilon^\mu P_\mu \delta_{ij},
\eeq
where $\epsilon=-2 i \bar {\hat \zeta} \Gamma^\mu \hat \xi$ and 
$P_\mu$ is the constant shift of $\hat \Phi$.

\section{Conclusions and discussion}

We have studied the world volume theory of the pairs of D-brane and ghost D-brane,
especially D$(-1)$-brane and ghost D$(-1)$-brane
and have seen that the nonlinear symmetries, 
the supersymmetries and the translation symmetry, 
are unbroken in the model. This is consistent with the interpretation 
of the system as the closed string vacuum.
We can extend our study in this paper to BFSS matrix model, 
and other dimensional branes.
We can also consider the type I superstring and supergroup $OSp$.
Of course, our discussion in this paper is 
rather naive and need further study.
In particular the problem of the unitarity may be important.

The nonlinear supersymmetries in 
pairs of D9-brane and ghost D9-brane will be also unbroken
in the same way as D$(-1)$-branes.
In this case, we can consider the instanton 
only on the D9-branes and in the small instanton limit
what we have is physically equivalent to a D5-brane with nothing \cite{instanton}.
(If we put the same instanton also on the ghost D9-branes, we have closed string vacuum
and if we put the anti-instanton on the ghost D9-branes, we have D5-anti-D5-brane pair.)
Then the half of the nonlinear supersymmetries generated by $\hat \zeta'$
which satisfies $\hat F_{\mu \nu} \Gamma^{\mu \nu} \hat \zeta'=0$ 
are expected to be unbroken, but others are broken from
$\langle \delta(\hat F_{\mu \nu} \Gamma^{\mu \nu} \hat \psi ) \rangle =0$.
Of course, the half of the linear supersymmetries generated by $\hat \zeta$
which satisfies $\hat F_{\mu \nu} \Gamma^{\mu \nu} \hat \zeta=0$
are also unbroken.
Thus we have different unbroken 16 supersymmetries from what D9-brane has
and if we consider the anti-instanton instead of the instanton we will have
the other half of unbroken supersymmetries.
This is interesting \cite{TeUe} since 
we discussed the 32 supersymmetries 
even though
for gauge theories without gravity 
16 supersymmetries are maximal in a usual sense.
However, there is a problem for this picture.
For the D9-brane-ghost D9-brane case,
the superalgebra is
\beq
[\delta_{\hat \zeta}, \delta_{\hat \xi}] =  \delta_{\epsilon}
+gauge \, transformation, \;\;\;
[\delta_{\hat \zeta}, \delta'_{\hat \xi'}] = \delta_{\epsilon'}, \;\;\; 
[ \delta'_{ {\hat \zeta'} }, \delta'_{\hat \xi'} ] = 0, 
\eeq
where $\delta_{\epsilon}$ is the translation by 
$\epsilon^\mu= 2 \hat \zeta \Gamma^\mu \hat \xi$.
In the compactified space-time,
$\delta_{\epsilon'}$ is the constant shift of $\hat A_\mu$, i.e. Wilson line, by 
$\epsilon^\mu= 2 \hat \zeta \Gamma^\mu \hat \xi'$ and 
corresponds to the space-time translation in the T-dual picture.
Thus for the uncompactified space-time,
$\delta_{\hat \zeta}$ 
%and $\delta_{\hat \zeta}-\delta_{\hat \zeta'}$  
form the 16 supersymmetries, but $\delta'_{\hat \zeta'}$ is trivial.
What we expect is that the $\delta'_{\hat \zeta'}$ also form the superalgebra.
To make this clear is an interesting question.

Another question is how to realize the Lorentz symmetry of
pairs of D$p$-brane and ghost D$p$-brane.
Since it mixes the coordinate and the fields in general,
to find the unbroken symmetry will be interesting.

\vskip6mm
\noindent
{\bf Acknowledgements}

\vskip2mm
We would like to thank 
T. Takayanagi
for useful discussions and comments.
This work was supported in part by DOE grant
DE-FG02-96ER40949.
%The works of 
%were supported in part by JSPS Research Fellowships for Young 
%Scientists. 
\\

\noindent
{\bf Note added}: 

As this article was being completed,
we became aware of the preprint \cite{Kawai}
%which partly overlaps the present work.
in which the matrix model proposed in this paper is also proposed.

%\appendix
%\setcounter{equation}{0}

\end{document}